\documentclass[aps,pra,twocolumn,10pt]{revtex4}
 \usepackage{amsthm}
\usepackage{graphicx,natbib,amssymb}
\usepackage{newlfont}
\usepackage{graphicx}
\usepackage{color}
\usepackage{subfigure}
\usepackage[section]{placeins}
\usepackage{psfrag}
\usepackage{caption2}
\usepackage{subeqn}
\usepackage{flafter}
\usepackage{bm}%
\begin{document}
\title{Stabilization of light bullets in nonlinear metamaterial waveguides}
\author{A. K. Shafeeque Ali}
\affiliation{\normalsize \noindent Department of Nonlinear Dynamics, School of Physics,
Bharathidasan University, Tiruchirappalli 620 024, India}
\author{A. Govindarajan}
\affiliation{\normalsize \noindent Department of Nonlinear Dynamics, School of Physics,
Bharathidasan University, Tiruchirappalli 620 024, India}
\author{M. Lakshmanan}
\affiliation{\normalsize \noindent Department of Nonlinear Dynamics, School of Physics,
Bharathidasan University, Tiruchirappalli 620 024, India}
\begin{abstract}
In this paper, we carry out a theoretical investigation on the propagation of spatio-temporal solitons (light bullets) in the nonlinear metamaterial waveguides. Our theoretical study is based on the formulation of Lagrangian variational analysis with a suitable ansatz followed by a split-step Fourier method in confirming the outcomes former numerically. A particular emphasis is given to obtain the conditions on the system parameters for stable dynamics in negative as well as positive index regimes of metamaterial waveguides. Similar to the conventional medium, the three-dimensional (3D) light bullets are highly unstable in metamaterials with the Kerr type nonlinearity alone. However, in the negative index regime of metamaterials, stable propagation of light bullets may occur in the normal dispersion regime balancing with defocusing cubic nonlinearity and focusing quintic nonlinearity. As in the conventional case, the stable dynamics is also observed in the case of anomalous dispersion with focusing cubic nonlinearity and defocusing quintic nonlinearity in the positive index regime. To test the solitonic nature of the 3D light bullets in the metamaterials, we also numerically investigate the collision dynamics of two light bullets. The study shows that the spatio-temporal soliton propagates without any change except perhaps some phase shift after a collision with another spatio-temporal soliton in competing cubic and quintic nonlinear metamaterials. The improper balancing between the linear and nonlinear effects results to form the bullet molecules in a distorted form with a large amount of energy after interaction and in the long run, oscillations of the light bullets grow and the bullets become filaments.  We have observed the same collision dynamics in both the negative and positive refractive index regimes of the metamaterial.
\end{abstract}
\maketitle
\section{Introduction}
Man-made materials with the negative real part of refractive index can display electromagnetic properties that are impossible to realize with positive index materials existing in nature. In the late 1960s, V. G. Veselago theoretically analyzed the light wave propagation in such an artificial material \cite{veselago}. This material, which consists of designed inclusions and has simultaneous negative values of electric permittivity and magnetic permeability, is commonly known as negative index material (NIM). In addition to the negative refraction, the NIM exhibits unusual characteristics such as reversed Doppler shift, reversed Goos-Hanchen shift, reversed Cerenkov radiation and reversal of Fermat's principle \cite{Ramakrishna, Caloz}. A design has been proposed to control the electric and magnetic responses by tuning the geometrical parameters of the thin wire lattice and split-ring resonator, which was the first remarkable suggestion to make the NIM practically possible  \cite{Pendry1, Pendry2, Pendry3}.  Following the above idea, Smith \emph{et al.} have realized a NIM experimentally using the thin wire lattice and split ring resonator as basic constituents \cite{Smith}. The metamaterials can also show nonlinear effects such as harmonic generations \cite{roppo}, which are designed by inserting nonlinear elements such as diodes \cite{Ilya} to the split-ring resonators or by embedding an array of meta-atoms into a nonlinear dielectric medium \cite{Zharov}.
\par
In recent times, investigation on the nonlinear pulse propagation in NIMs is being actively pursued. The nonlinear partial differential equation governing the propagation of light pulse in the NIM has been derived and is found to admit envelope solitary wave solutions \cite{Partha}. A generalized nonlinear Schr\"{o}dinger
equation for dispersive dielectric susceptibility and permeability has been derived to
describe the propagation of electromagnetic pulse in the NIM and it has been found that the
linear properties of the medium can be tuned to modify its linear as well as nonlinear effective properties, which lead to a new form of dynamical behavior \cite{Maxim}. The role of second-order nonlinear dispersion in the stable propagation  of Gaussian pulse in the NIM in focusing or defocusing cases with normal
or anomalous regimes has been analyzed \cite{Ancemma}. The evolution equations for the envelopes of beams and spatio-temporal pulses in nonlinear dispersive NIM have been derived and stability of solitary wave solutions has been analyzed using numerical methods based on fast Fourier$-$Bessel transforms \cite{GN}.  By adopting the methods of quantum statistics and a kinetic equation for the pulses the partial coherence in NIMs has been discussed \cite{GN1}. The Raman soliton self-frequency shift in the nonlinear NIMs can be controlled by nonlinear electric polarization \cite{GN2}. The existence of gray solitary waves and the conditions for their formation in NIMs have also been studied \cite{new11}. The propagation of ultrashort electromagnetic pulse in metamaterials with cubic electric and magnetic nonlinearities have been investigated and it has been predicted that spatio-temporal electromagnetic solitons may exist in the negative index regime of the metamaterials with defocusing nonlinearity and normal group velocity dispersion \cite{wen1}. The self-focusing of ultrashort pulses in NIMs can be controlled by tailoring dispersive magnetic permeability \cite{Jinggui}. In contrast to ordinary positive index materials, in the case of NIM  dark solitons may exist for the case of normal second-order dispersion, anomalous third-order dispersion, self-focusing Kerr nonlinearity, and non-Kerr nonlinearities \cite{Vivek}. Modulational instability and the generation of ultrashort pulses in NIMs have also been investigated in Refs. \cite{modu1,modu2,modu3,wen2,modu4,modu5}.
\par
It is well known that optical solitons in a medium with Kerr optical nonlinearity are unstable in two and three dimensions due to the phenomenon of optical beam collapse \cite{bu1}. However, several investigations have been carried out to make intrinsic wave collapse-free light bullets. Stabilization of spatio-temporal light bullets can be achieved in the presence of quadratic nonlinearity \cite{bu2}, nonlocal nonlinear response \cite{bu3}, negative fourth-order dispersion \cite{bu31}, higher-order nonlinear media \cite{bu4, bu41}, cylindrical Bessel lattice \cite{bu5}, semi-infinite array of weakly coupled nonlinear optical waveguides \cite{bu51}, filamentation of femtosecond pulses \cite{bu6}, harmonic and parity-time-symmetric potentials \cite{bu7} and so on. Among the possible ways to stabilize the light bullets, we opt for the higher order nonlinear media as there are quite a few materials which naturally support higher order nonlinear dielectric susceptibilities, including $CdS_xSe_{1-x}$ doped glasses \cite{acioli}, $AlGaAs$ semiconductors \cite{tanev} and chalcogenide glass \cite{lawrence, vigneshraja2019,Tamil1,raja2020tailoring} which can even support upto septimal nonlinearity when the pulse is injected with a moderate peak power. Motivated by these facts, in the present investigation, we carry out a theoretical study on the stable dynamics of light bullets in metamaterial waveguides with competing cubic and quintic nonlinearities. As pointed above, metamaterials are engineered materials and their electromagnetic properties can be tuned at will. Enjoying the engineering freedom to alter electromagnetic properties, here we will arrive at special conditions to achieve stable propagation of light bullets in metamaterials. We consider a nonlinear Schr\"{o}dinger type equation with higher-order nonlinearities as the propagation model and investigations are carried out by adopting the Lagrangian variational method and numerical analysis. We find that the stability of light bullets in the negative index regime of metamaterials with the normal dispersion regime can be enhanced by the combined action of defocusing cubic nonlinearity and positive quintic nonlinearity. On the other hand, the influence of focusing cubic nonlinearity and negative quintic nonlinearity make the light bullets propagate stable in the positive index regime. We also show that the 3D light bullet propagates without any change after a collision with another light bullet in competing cubic and quintic nonlinear metamaterial waveguides.
\par
The rest of the paper is organized as follows. In section II the theoretical model of the problem is presented and the stability criterion for the propagation of three-dimensional light bullets is obtained by adopting the Lagrangian variational method. Numerical investigations on the propagation of spatio-temporal solitons and their collision dynamics in metamaterial waveguides have been carried out in detail in section III. Conclusions are made in section VI.
\section{Theoretical model and variational formulation}
The governing model that describes the propagation of electromagnetic waves in negative index materials is given by the following nonlinear
Schr\"{o}dinger equation \cite{wen1, wen2},
\begin{equation}
\frac{\partial \xi}{\partial z }=-i\,\frac{sgn(\beta_2)}{2}\, \frac{\partial^2 \xi}{\partial t^2}+ i\,\frac{sgn(n)}{2}\nabla^2_\bot \xi+i \gamma |\xi|^{2}\,\xi+i \vartheta |\xi|^{4}\,\xi,
\label{modeleqn11}
\end{equation}
where $\xi(x, y, z, t)$ represents the normalized complex amplitude of the
propagating modes. sgn$(\beta_2)$ = $\pm1$ corresponds to normal and anomalous
group velocity dispersions, respectively, and sgn(n) = $\pm1$ corresponds to
positive and negative index of refractions, respectively. $\nabla^2_\bot=\frac{\partial^2}{\partial x^2}+\frac{\partial^2}{\partial y^2}$ is the transverse Laplace operator. The terms, $z$ and $t$ stand for the direction of propagation of the beam and the time in a co-moving frame of reference, respectively. Also, $\gamma$ and $\vartheta$ stand for cubic and quintic nonlinear coefficients, respectively.
Now, we will utilize the Lagrangian variational method \cite{Anderson, Anderson1} involving a trial function to study the dynamics of the light bullets in the metamaterial waveguides. The basic idea of the method is to make use of the Lagrangian density to identify an effective Lagrangian for the system under the consideration.  The associated Lagrangian for the propagation
model of negative index materials as given in
Eq. (\ref{modeleqn11}) is expressed by \cite{Aks1}
\begin{eqnarray}
L= \frac{i}{2}\,(\xi \frac{\partial \xi^*}{\partial z }-\xi^*\frac{\partial \xi}{\partial z })-\,\frac{sgn(\beta_2)}{2}  |\frac{\partial \xi}{\partial t }|^2 \nonumber \\ +\,\frac{sgn(n)}{2}  |\frac{\partial \xi}{\partial x }|^2 +\,\frac{sgn(n)}{2}  |\frac{\partial \xi}{\partial y }|^2-\frac{\gamma}{2}\,|\xi|^{4}-\frac{\vartheta}{3}\,|\xi|^{6}.
\end{eqnarray}
We consider the Gaussian ansatz of the following form \cite{konar},
\begin{eqnarray}
\xi(z,x,y,t)=\Phi(z)e^{i\theta(z)} e^{-\frac{(x^2+y^2+t^2)}{2 a(z)^2}} e^{i\alpha(z)( x^2+ y^2+t^2)},
\end{eqnarray}
where $\Phi(z)$ is the amplitude and $a(z)$ is the width. It is worth noting that we have here studied the spatio-temporal solitons (light bullets) with a spherically symmetric structure and hence we kept the width equal in the space and time domains. Though one can also study the light bullets with elliptical structure by keeping different widths in the space and time \cite{mihalache,boris1}, the light bullets with equal width are the typical candidates indicating solitonic nature and predominantly explored in experimental settings too. Here, 
$\alpha(z)$ is the parameter to account for the
chirp. The reduced Lagrangian of
the system can be calculated using the following equation,
\begin{equation}
\langle\pounds\rangle=\int_{-\infty}^\infty \int_{-\infty}^\infty \int_{-\infty}^\infty L dx dy dt.
\label{leff}
\end{equation}
Consequently, the effective Lagrangian becomes,
\begin{eqnarray}
\label{leffeq}
\langle\pounds\rangle=\pi^{3/2}\Phi(z)^2\{\frac{a(z)}{4}[2sgn(n)-sgn(\beta_2)]+a(z)^5 \nonumber \\  \alpha(z)^2[2sgn(n)-sgn(\beta_2)]-\frac{\gamma }{4 \sqrt{2}} a(z)^3\Phi(z)^2\nonumber \\  -\frac{\vartheta }{9 \sqrt{3}} a(z)^3\Phi(z)^4+a(z)^3[\frac{3}{2}\frac{\partial \alpha(z)}{\partial z } a(z)^2+\frac{\partial \theta(z)}{\partial z}]\}.
\end{eqnarray}
Now, we vary Eq. (\ref{leffeq}) with respect to different beam parameters $\Phi(z)$, $\theta(z)$, $a(z)$ and $\alpha(z)$, which yield the following evolution equations:
\begin{figure}[!h]
\begin{center}
\includegraphics*[height=5cm, width = 7cm]{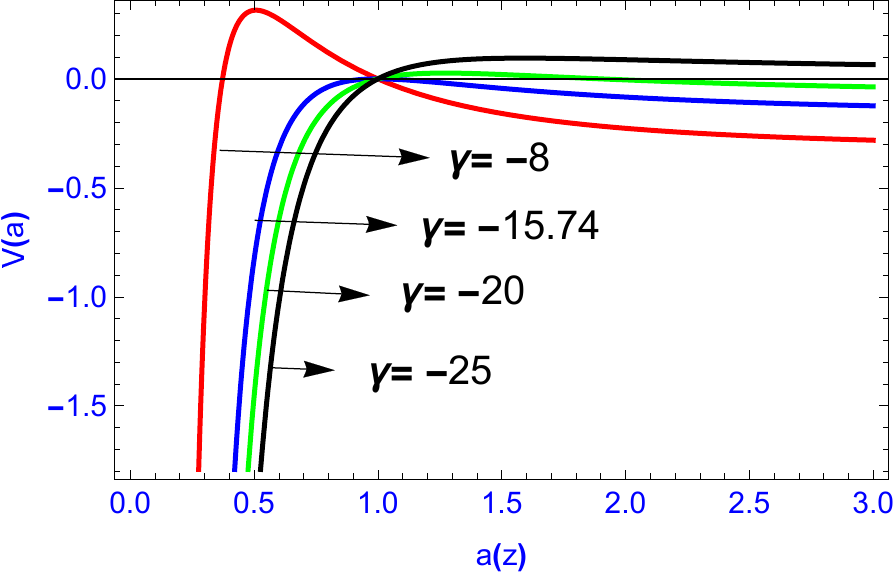}
\caption {(Color online) Potential function ($V(a)$) versus width ($a(z)$) as a function of cubic nonlinearity ($\gamma$) in the negative index materials. Other parameters are $sgn(\beta_2)=1$, $P=1$ and $\vartheta=0$.}
\label{pcr1}
\end{center}
\end{figure}\\
(i): combining the variations of $\theta(z)$ and $\alpha(z)$:
\begin{equation}
\label{6a}
\alpha(z)=\frac{3}{2}\frac{1}{[2 sgn(n)-sgn(\beta_2)] a(z)}\frac{\partial a(z)}{\partial z},
\end{equation}
(ii): variation of $\theta(z)$:
\begin{eqnarray}
\label{6b}
P=\pi^{3/2}a(z)^3 \Phi(z)^2,
\end{eqnarray}
(iii): combining the variations of $a(z)$ and $\Phi(z)$:
\begin{eqnarray}
\label{6c}
\frac{\partial^2 a(z)}{\partial z^2}=\frac{1}{9 a(z)^3}[5-4 sgn(n)sgn(\beta_2)] \nonumber \\-\frac{\gamma \Phi(z)^2}{6 \sqrt{2} a(z)}[2 sgn(n) - sgn(\beta_2)] \nonumber\\-\frac{\vartheta \Phi(z)^4 }{27 \sqrt{3} a(z)}[8 sgn(n) -4  sgn(\beta_2)],
\end{eqnarray}
Eq. (\ref{6b}) shows that $\pi^{3/2}a(z)^3 \Phi(z)^2$ is a conserved quantity throughout the propagation of the beam as the corresponding conjugate momenta is conserved in Eq. (\ref{leffeq}). Using Eq. (\ref{6b}), Eq. (\ref{6c}) can be rewritten as,
\begin{eqnarray}
\label{M6b1}
\frac{\partial^2 a(z)}{\partial z^2}=\frac{1}{9 a(z)^3}[5-4 sgn(n)sgn(\beta_2)] \nonumber \\-\frac{\gamma P}{6 \sqrt{2} \pi^{3/2}a(z)^4}[2 sgn(n) - sgn(\beta_2)] \nonumber\\-\frac{\vartheta P^2 }{27 \sqrt{3} \pi^{3} a(z)^7}[8 sgn(n) -4  sgn(\beta_2)].
\end{eqnarray}
Integrating Eq. (\ref{M6b1}) once, one can obtain the potential-well description as follows
\begin{equation}
\frac{1}{2}(\frac{\partial a(z)}{\partial z })^2+V(a)=0,
\label{poten}
\end{equation}
\begin{figure}
    \subfigure[]{\label{new2}\includegraphics[height=5 cm, width=7 cm]{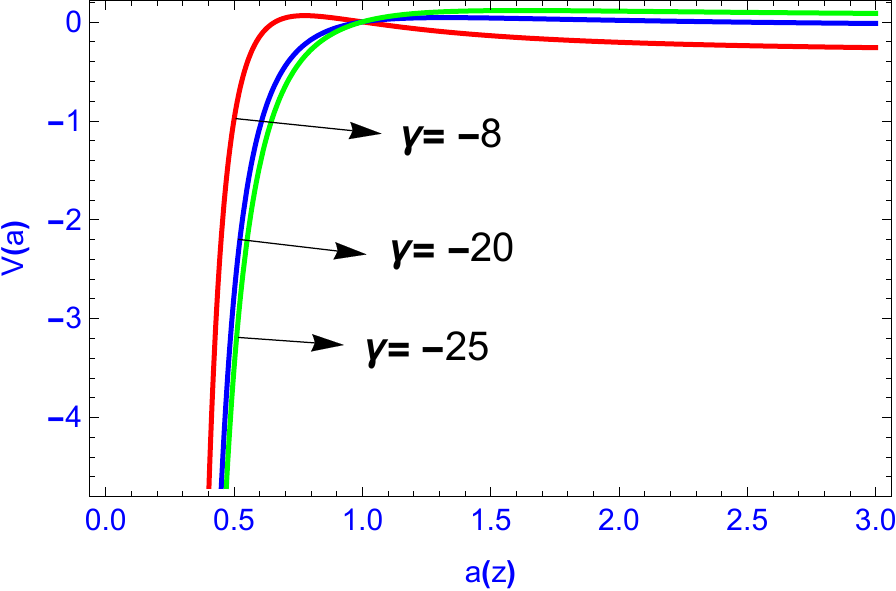}}
    \subfigure[]{\label{new221}\includegraphics[height=5 cm, width=7 cm]{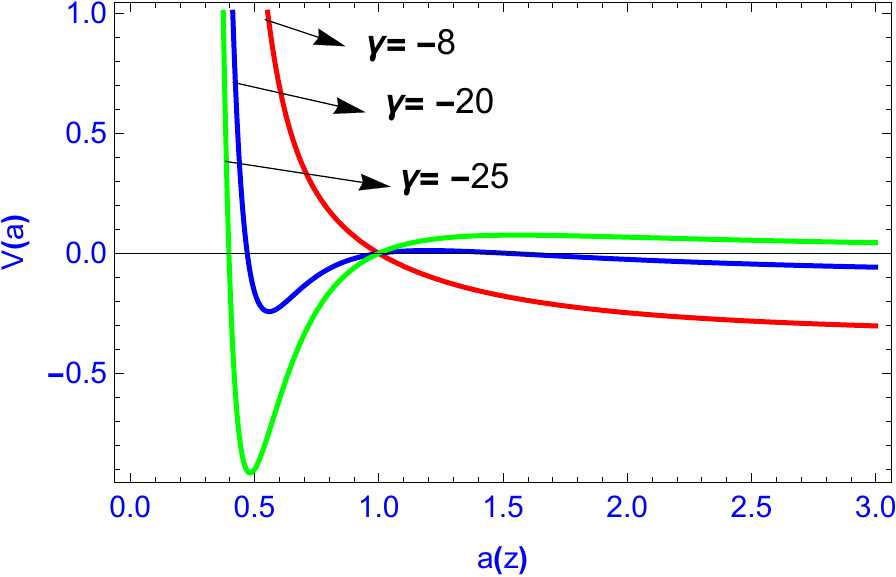}}
    \subfigure[]{\label{new222}\includegraphics[height=5 cm, width=7 cm]{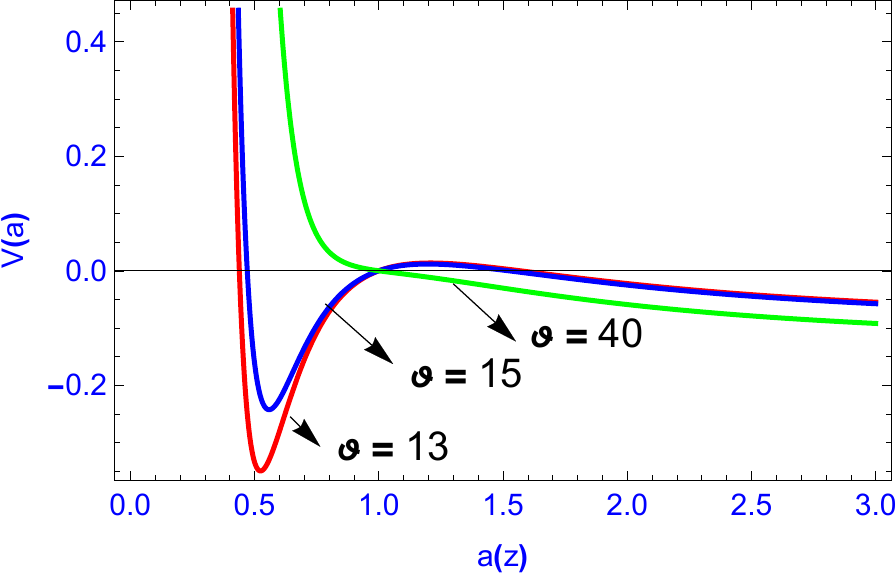}}
\caption{(Color online) Potential function ($V(a)$) versus width ($a(z)$) for different values of cubic and quintic nonlinear coefficients in the negative index materials with (a) $\vartheta=-15$, (b) $\vartheta=15$ and (c) $\gamma=-20$. Other parameters are $sgn(\beta_2)=1$, $sgn(n)=-1$ and $P=1$. }
  \label{m1}
\end{figure}
\begin{figure*}
\begin{center}
\subfigure[]{\label{a1in}\includegraphics[height=3.5 cm, width=4.5 cm]{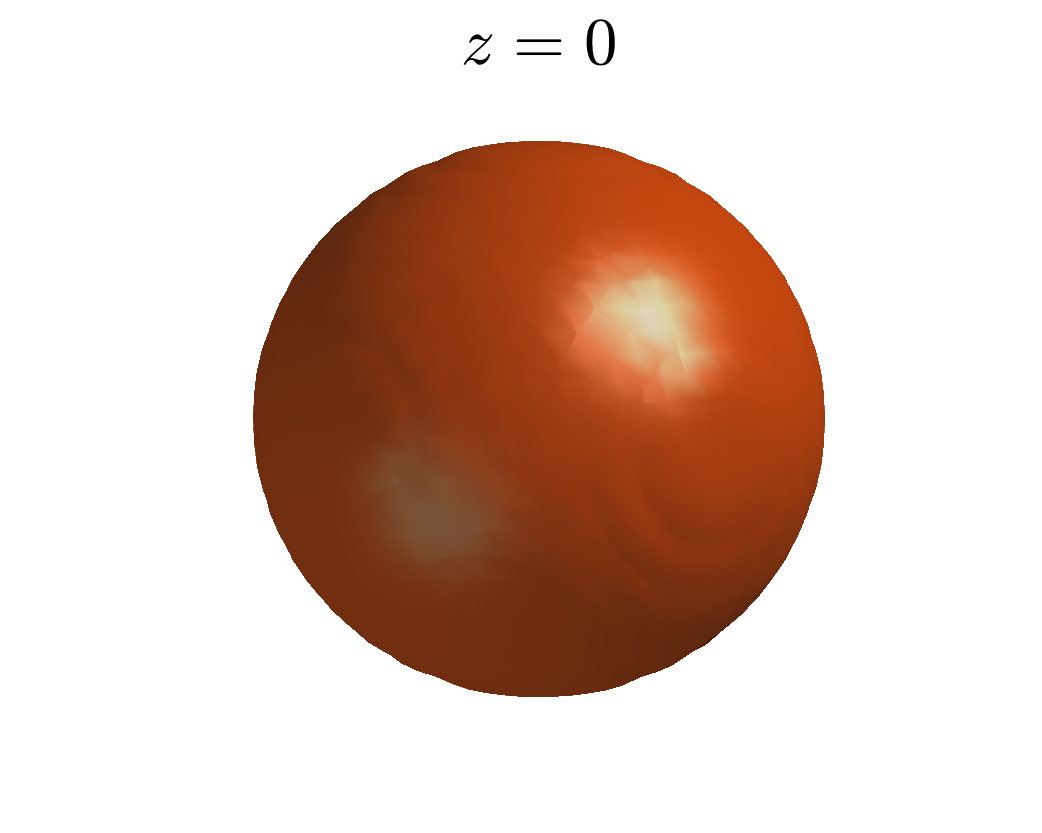}}
\subfigure[]{\label{a11}\includegraphics[height=3.5 cm, width=4.5 cm]{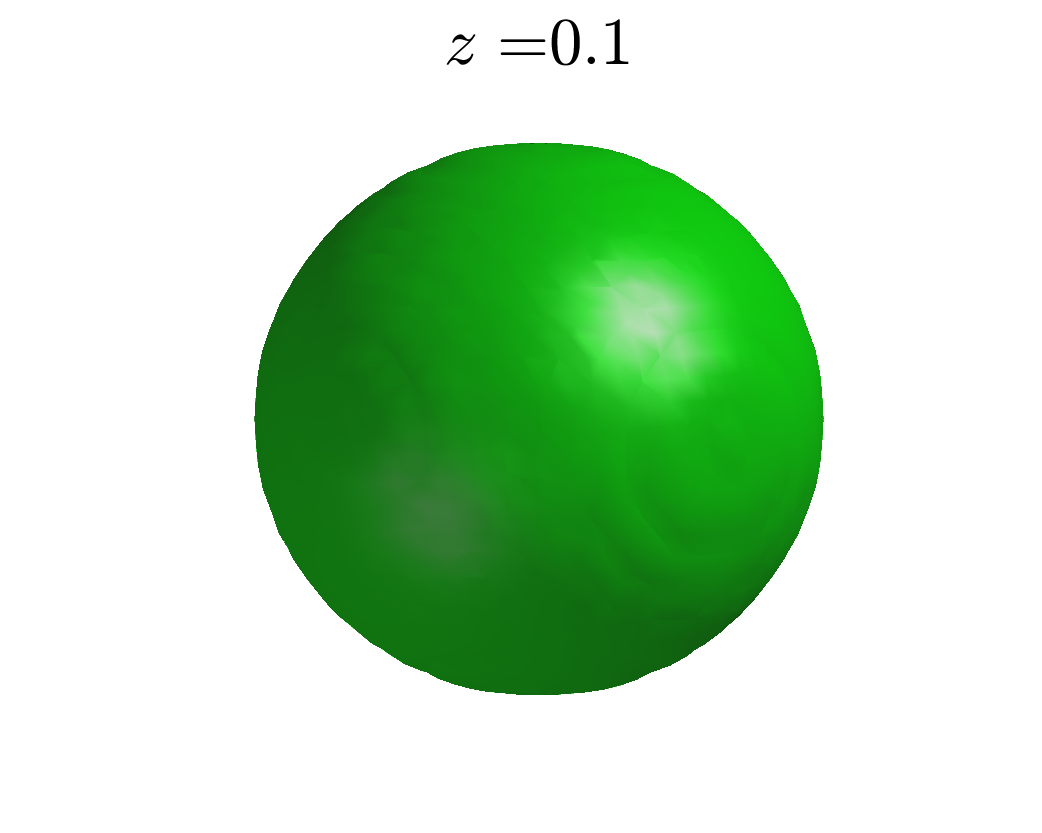}}
\subfigure[]{\label{a12}\includegraphics[height=3.5 cm, width=4.5 cm]{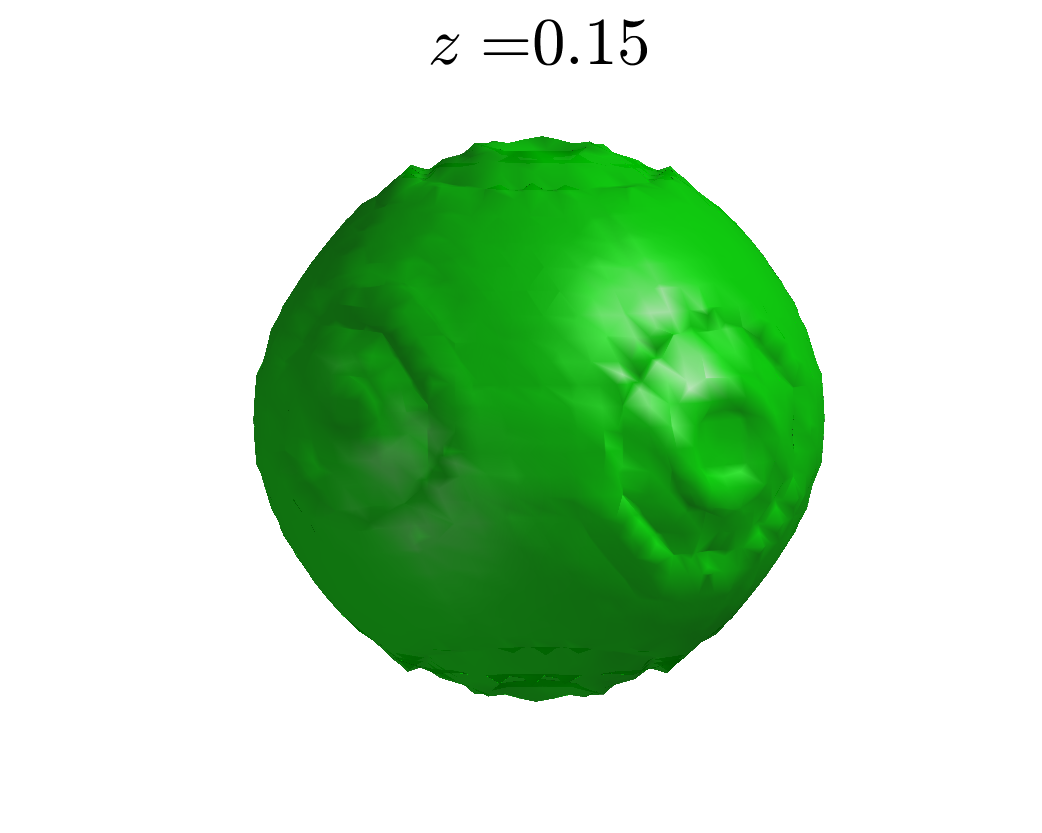}}
\subfigure[]{\label{a13}\includegraphics[height=3.5 cm, width=4.5 cm]{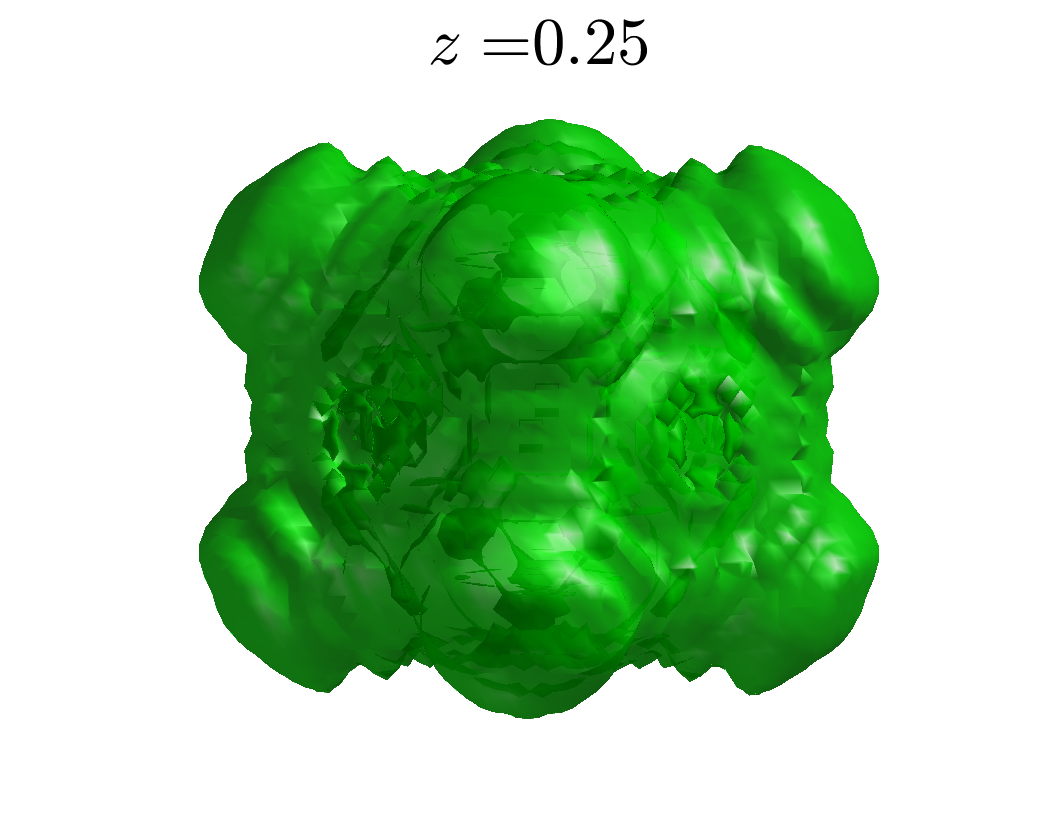}}
\subfigure[]{\label{a14}\includegraphics[height=3.5 cm, width=4.5 cm]{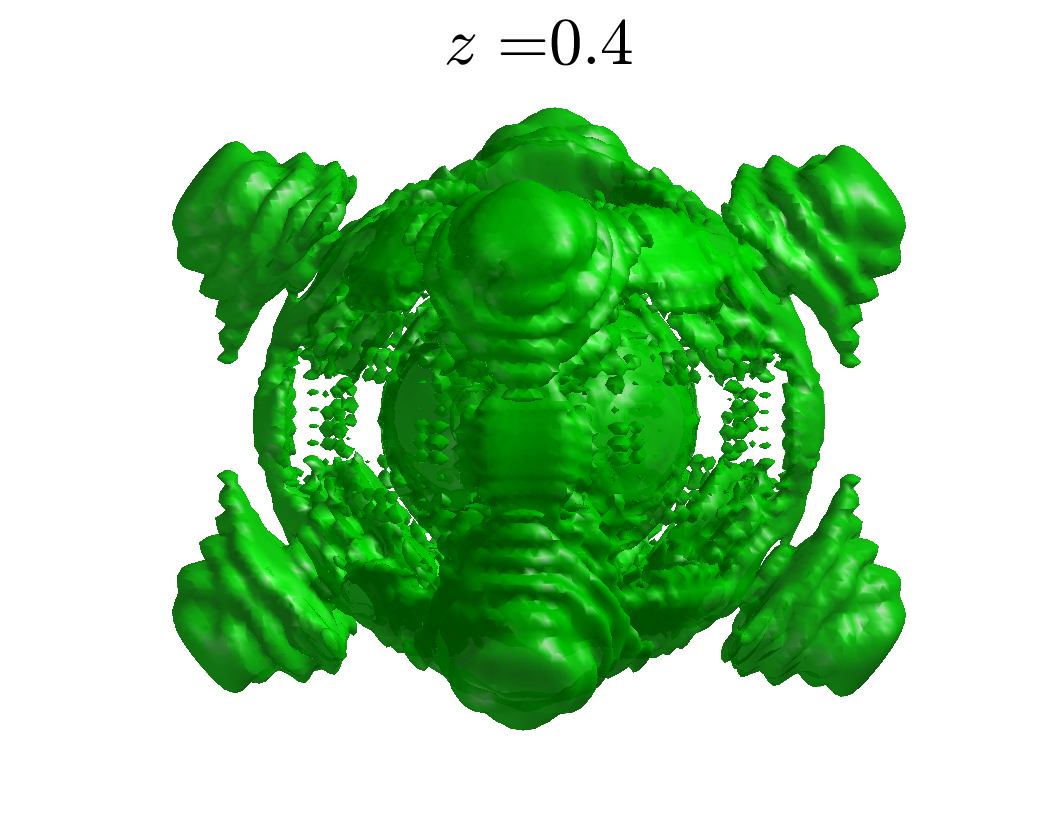}}
\subfigure[]{\label{a15}\includegraphics[height=3.5 cm, width=4.5 cm]{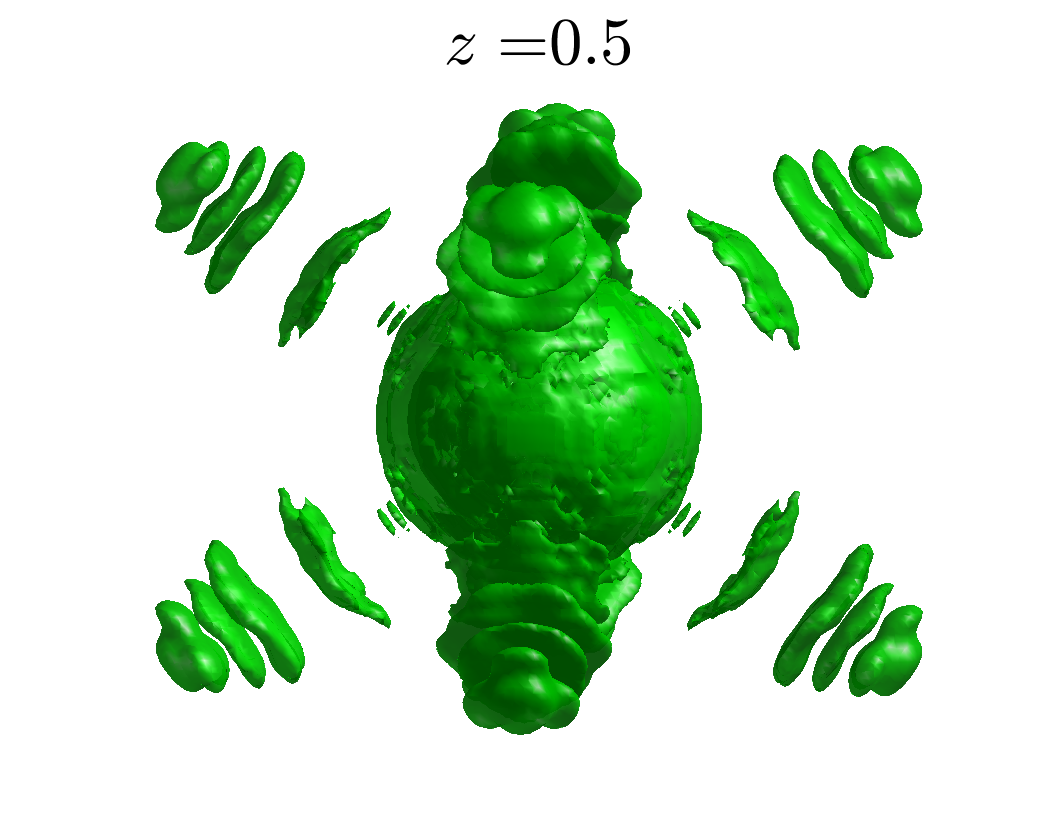}}
\caption{(Color online) Iso-surface plots depicting the evolution of 3D light bullets corresponding to the collapse in the negative index regime of metamaterial for P=1 with cubic nonlinearity alone. Other parameters are $sgn(\beta_2)=1$, $\gamma$=-20 and $\vartheta$=0.}
\label{a1}
\end{center}
\end{figure*}
 where the potential V(a) is given by
\begin{eqnarray}
\label{Vqq}
V(a)=(\frac{1}{ a(z)^2}-1)[\frac{5-4 sgn(n)sgn(\beta_2)}{18}]\nonumber\\-\frac{\gamma P}{18 \sqrt{2} \pi^{3/2}}(\frac{1}{ a(z)^3}-1)[2 sgn(n) - sgn(\beta_2)]\nonumber\\-\frac{2\vartheta P^2}{81 \sqrt{3} \pi^{3}}(\frac{1}{ a(z)^6}-1)[2 sgn(n) - sgn(\beta_2)].
\end{eqnarray}
It is clear from Eq. (\ref{Vqq}) that the stability of the electromagnetic wave propagating in the metamaterial waveguides depends on the sign of refraction, nature of dispersion and nonlinearity. If there is an adequate balance between the system parameters, then the spatio-temporal solitons propagate without suffering any distortion. In the metamaterials with Kerr type nonlinearity, the equilibrium point at the potential energy curve exists with a particular input power called critical power (see Appendix for details). Hence the light bullets may exist when the input power satisfies the following condition,
 \begin{equation}
 P_{c1}=\frac{2 \sqrt{2}\pi ^{3/2} a(z)}{3 \gamma} (2 sgn(n)-sgn(\beta_2)).
 \label{pow1}
\end{equation}
Let us assume the initial width, $a(0)=1$. In the case of negative (positive) index regime of the metamaterial with normal (anomalous) dispersion and defocusing (focusing) cubic nonlinearity the light bullets may exist when the input power, $ P=P_{c1}=\frac{2 \sqrt{2} \pi ^{3/2}}{|\gamma|}$, but will be unstable as the extremum is a maximum. The electromagnetic wave diverges when $ P<P_{c1}=\frac{2 \sqrt{2} \pi ^{3/2}}{|\gamma|}$ and  undergoes a collapse for $P>P_{c1}=\frac{2 \sqrt{2} \pi ^{3/2}}{|\gamma|}$. From the above relation, it is found that the critical power, $P_{c1}$  is inversely proportional to the strength of the cubic nonlinearity. The critical power and the equilibrium point at the potential energy curve are highly influenced by the value of cubic nonlinear coefficient. The dependence of the potential function  on the cubic nonlinearity in the negative index materials with input power $P=1$, $sgn(\beta_2)=1$ and $\vartheta=0$ has been depicted in Fig.\ref{pcr1}. In this investigation, we choose all the parameters in normalized units as the governing model given by Eq. (\ref{modeleqn11}) is a normalized equation. Here we intend to understand the parametric region which supports the stable dynamics of light bullets by comparing the system variables with each other. One
can see similar works with normalized units in the literature such as the optical bistability and gap soliton formation in the optical
metamaterial coupler \cite{NLI}, breather formation of the spatio-temporal vortex light bullets \cite{co1}, spatial ring formation in the nonlinear metamaterials \cite{wen1},  dynamics of higher-order solitary waves in the quadratic media \cite{DVS} and so on. It is quite clear from Fig. \ref{pcr1} that even though the input power $P =1$ and  the cubic nonlinear coefficient, $|\gamma|=2 \sqrt{2}\pi ^{3/2}=15.74$ ($P_{c1}$=P=1) the equilibrium point is not a stable one. Small deviations  from the equilibrium point will make the light bullets to be unstable and may lead to collapse or divergence.
\begin{figure*}
\begin{center}
\subfigure[]{\label{a2in}\includegraphics[height=4 cm, width=5 cm]{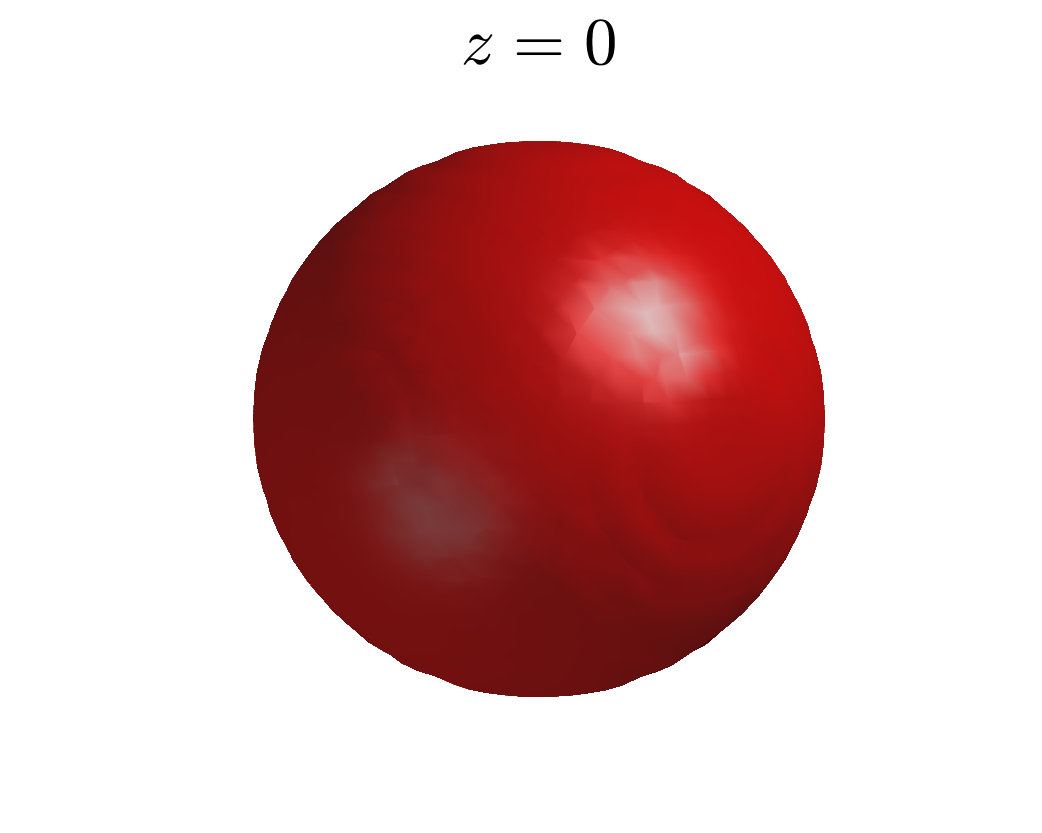}}
\subfigure[]{\label{a21}\includegraphics[height=4 cm, width=5 cm]{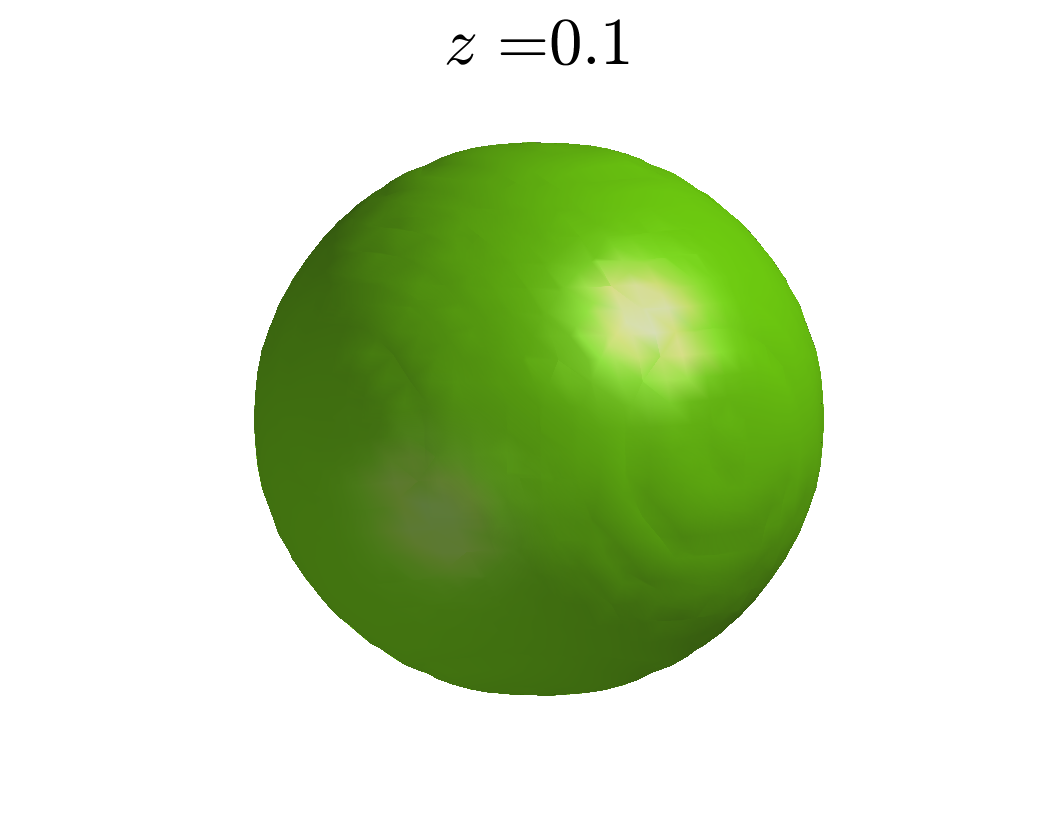}}
\subfigure[]{\label{a22}\includegraphics[height=4 cm, width=5 cm]{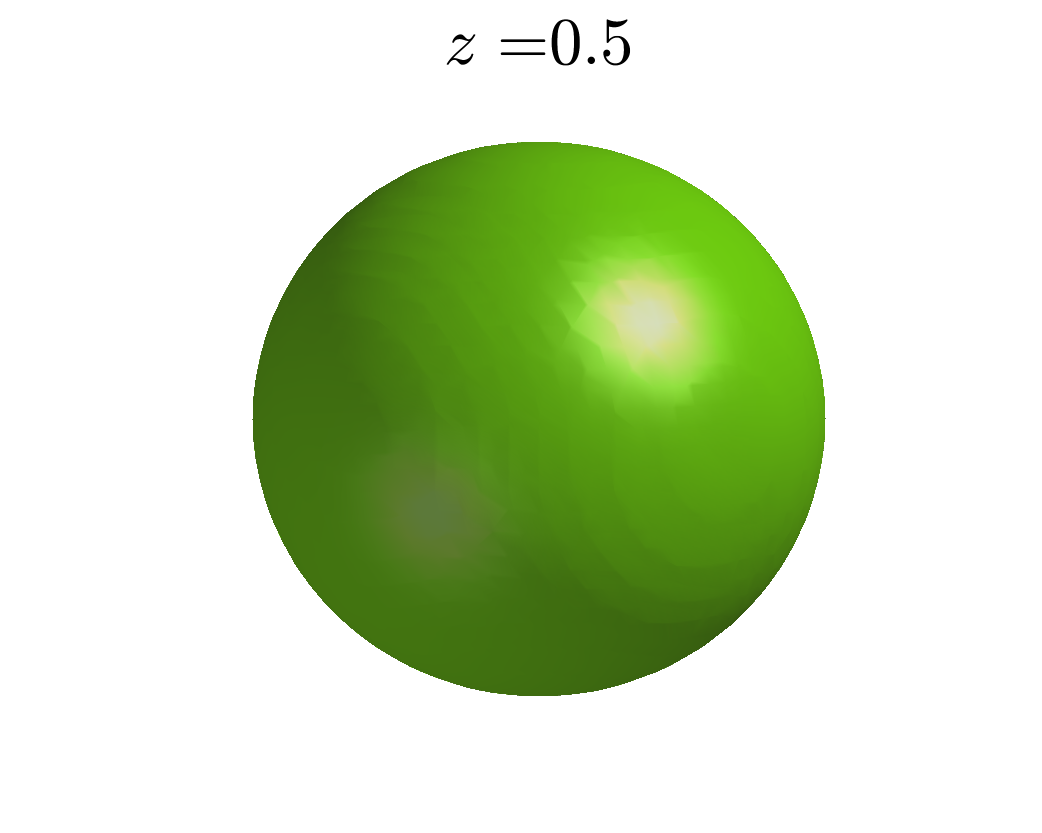}}
\subfigure[]{\label{a23}\includegraphics[height=4 cm, width=5 cm]{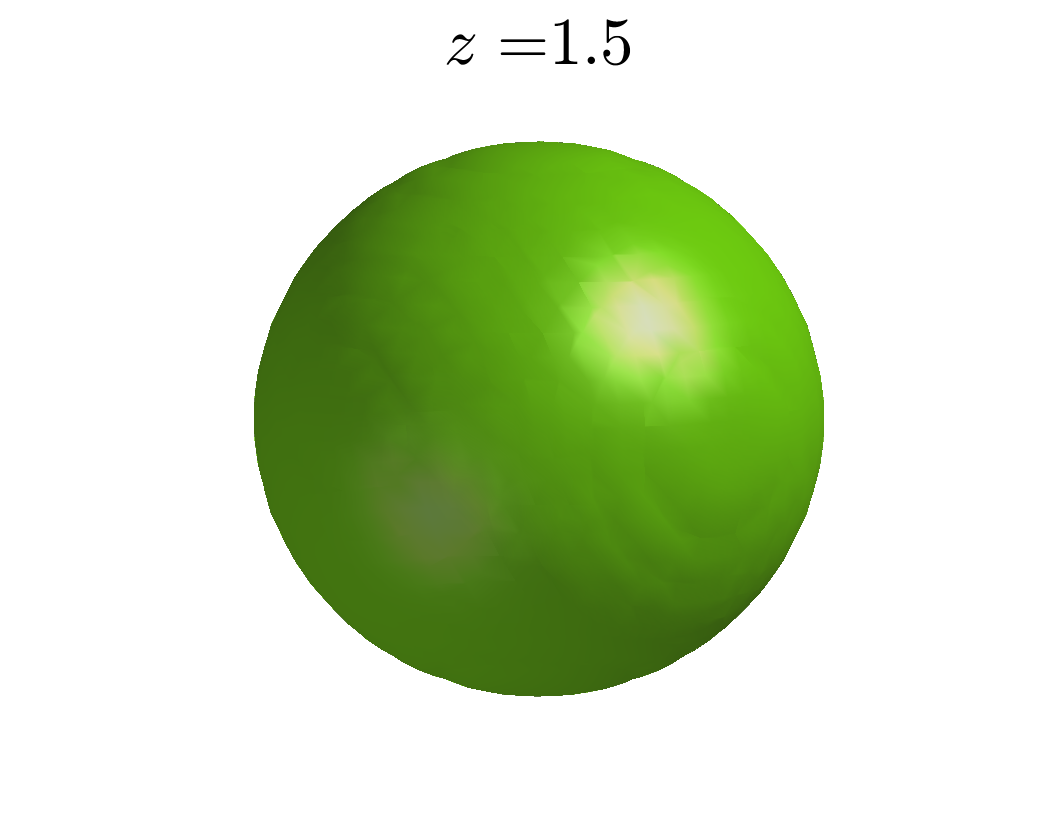}}
\subfigure[]{\label{a24}\includegraphics[height=4 cm, width=5 cm]{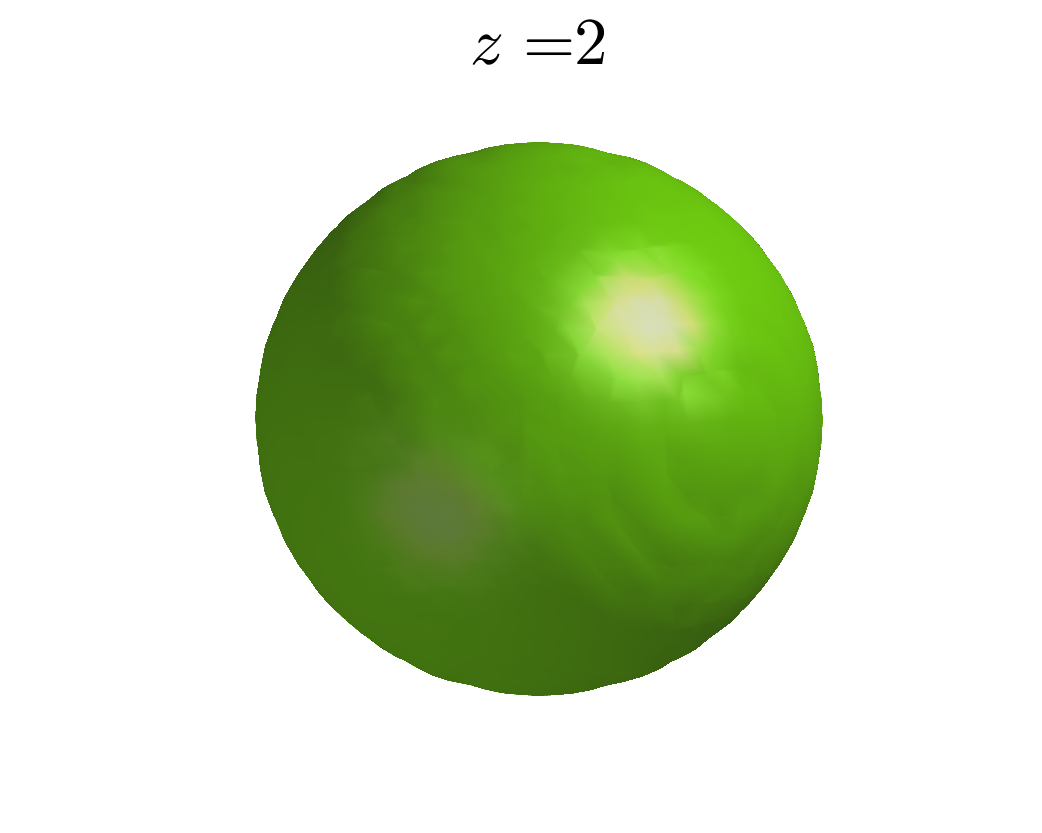}}
\caption{(Color online) Iso-surface plots indicating the evolution of stable 3D light bullets in the negative index regime of metamaterial for P=1 with competing cubic and quintic nonlinearities. Other parameters are $sgn(\beta_2)=1$, $\gamma$=-20 and $\vartheta$=15.}
\label{a2}
\end{center}
\end{figure*}
Now in the presence of quintic nonlinearity in addition to the cubic nonlinearity, the minimum critical power required by the system to support the formation of spatio- temporal solitons is given by the condition,
 \begin{eqnarray}
 P_{c2}=-\frac{9 \sqrt{\frac{3}{2}}a(z)^3 \gamma \pi ^{3/2}}{16 \vartheta}\pm\nonumber \\ \frac{\sqrt{1458 \gamma^2 \pi ^3 a(z)^6-2304 \pi ^3 \sqrt{3} \vartheta a(z)^4( sgn(\beta_2) -2 sgn(n))}}{32 \sqrt{3} \vartheta}.
 \label{pow2}
 \end{eqnarray}
When $P<P_{c2}$ the potential energy curve will not show any stable equilibrium point, but the energy minima and hence the stable bullets with negative energy can be observed if $P\geq P_{c2}$. It is clear from Eq. (\ref{pow2}) that the critical power $P_{c2}$ is a function of the cubic and quintic nonlinear coefficients. Proper tuning of these nonlinear coefficients is necessary to observe the stable dynamics of the light bullets.
\par
 Figure \ref{m1} depicts the relation between the potential function and width for different possible combinations of cubic and quintic nonlinearities in the negative index regime of metamaterials. Figure \ref{new2} shows the case where the dispersion is normal, with defocusing cubic and quintic nonlinearities. In this case, the propagation of the light bullet is unstable. Even though, there exist some equilibrium points, which are indeed unstable and minor fluctuations from these points may lead to the collapse or divergence of the light bullets. The stability of the spatio-temporal solitons is enhanced with the presence of competing cubic and quintic nonlinearities as depicted in Fig. \ref{new221}. The minimum potential energy in the figure corresponds to a stable spatio-temporal bullet with negative energy. In the case of the negative index regime, the potential well with energy minima and hence stable bullets with negative energy can be observed when quintic nonlinearity is of focusing type, whereas it is observed with defocusing quintic nonlinearity in the positive index regime of the metamaterials. Hence, in the negative index regime of the metamaterials, stable propagation of light bullets can occur in the normal dispersion regime balancing with defocusing cubic nonlinearity and focusing quintic nonlinearity whereas it is observed in the anomalous dispersion regime with focusing cubic nonlinearity and defocusing quintic nonlinearity in the positive index regime. The potential energy curve as a function of quintic nonlinearity for the case of $\gamma=-20$ is depicted in Fig. \ref{new222}. It is clear from the figure that the depth of the potential well decreases as the value of quintic nonlinearity increases. At a distinct value of input power, for a particular value of quintic nonlinearity, the potential well completely vanishes and the light bullets lose their stability nature. This is attributed to the fact that when the value of the quintic nonlinearity increases the value of the critical power $(P_c)$ at which the light bullet exists increases.
The Lagrangian variational method followed in this investigation is an approximation and it cannot predict the exact dynamics of light bullets in a medium. In order to study the exact evolution of the 3D bullets, we carry out a detailed numerical analysis in the following section.
 \section{Numerical Analysis}
 \begin{figure*}
\begin{center}
\subfigure[]{\label{coin}\includegraphics[height=4 cm, width=5 cm]{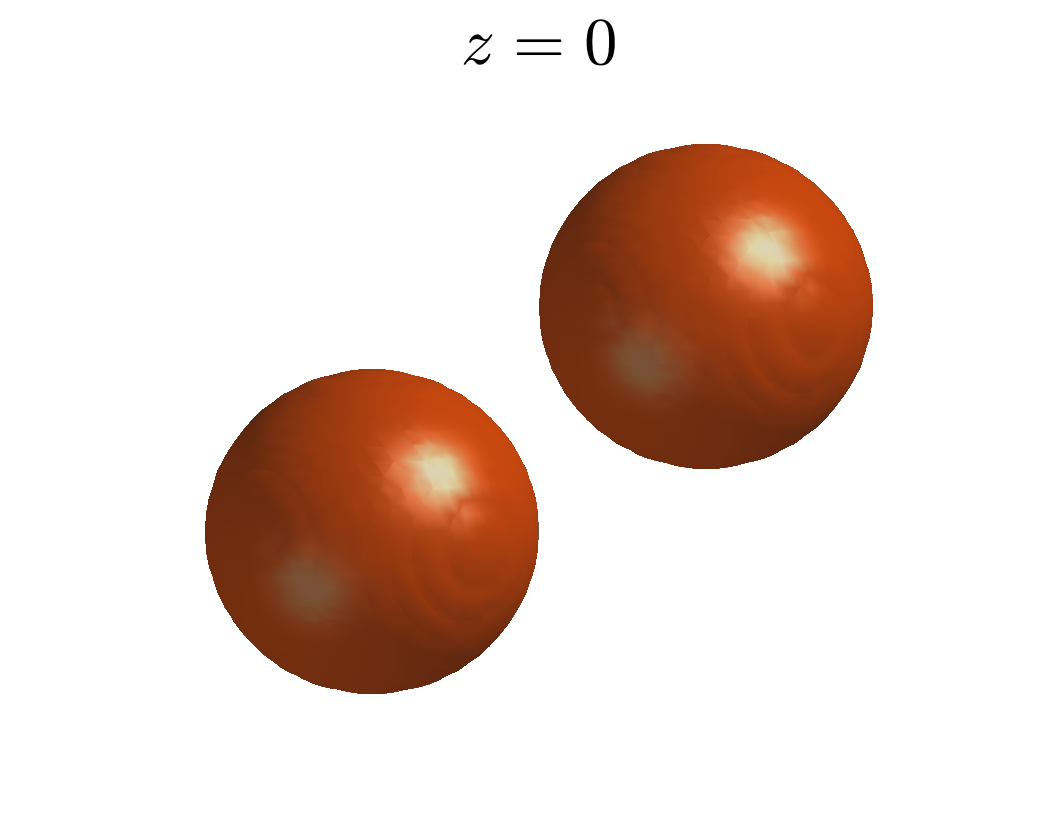}}
\subfigure[]{\label{co1}\includegraphics[height=4 cm, width=5 cm]{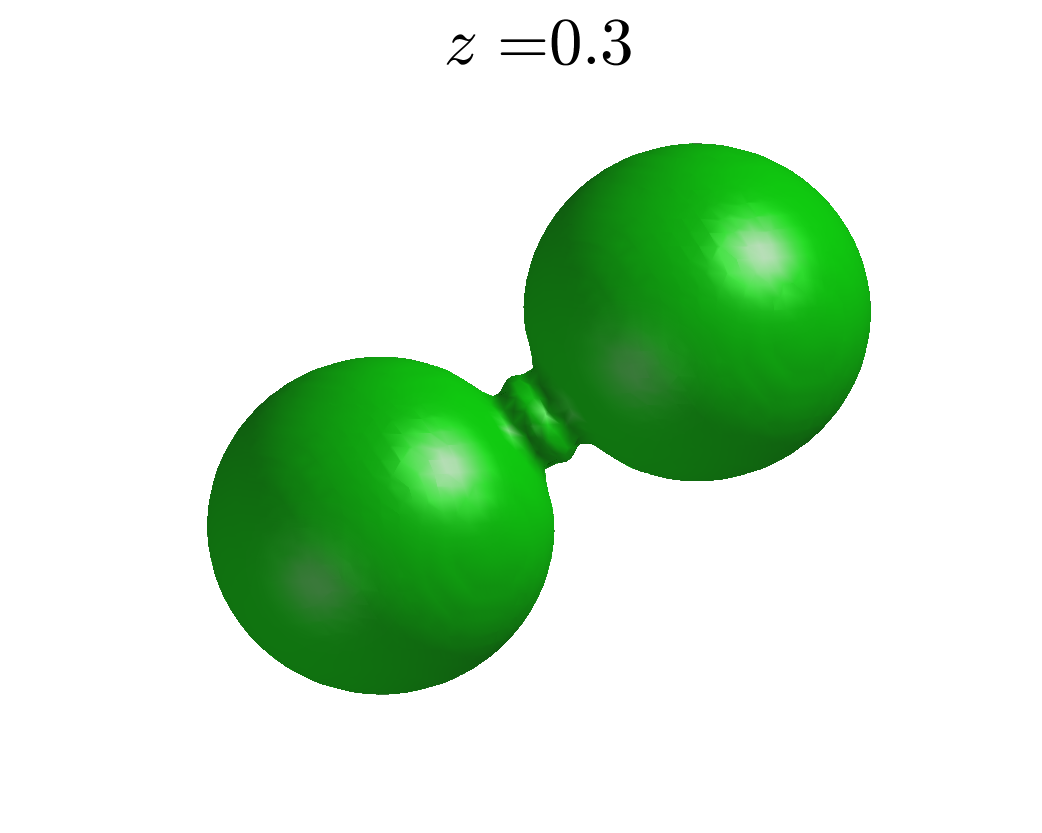}}
\subfigure[]{\label{co2}\includegraphics[height=4 cm, width=5 cm]{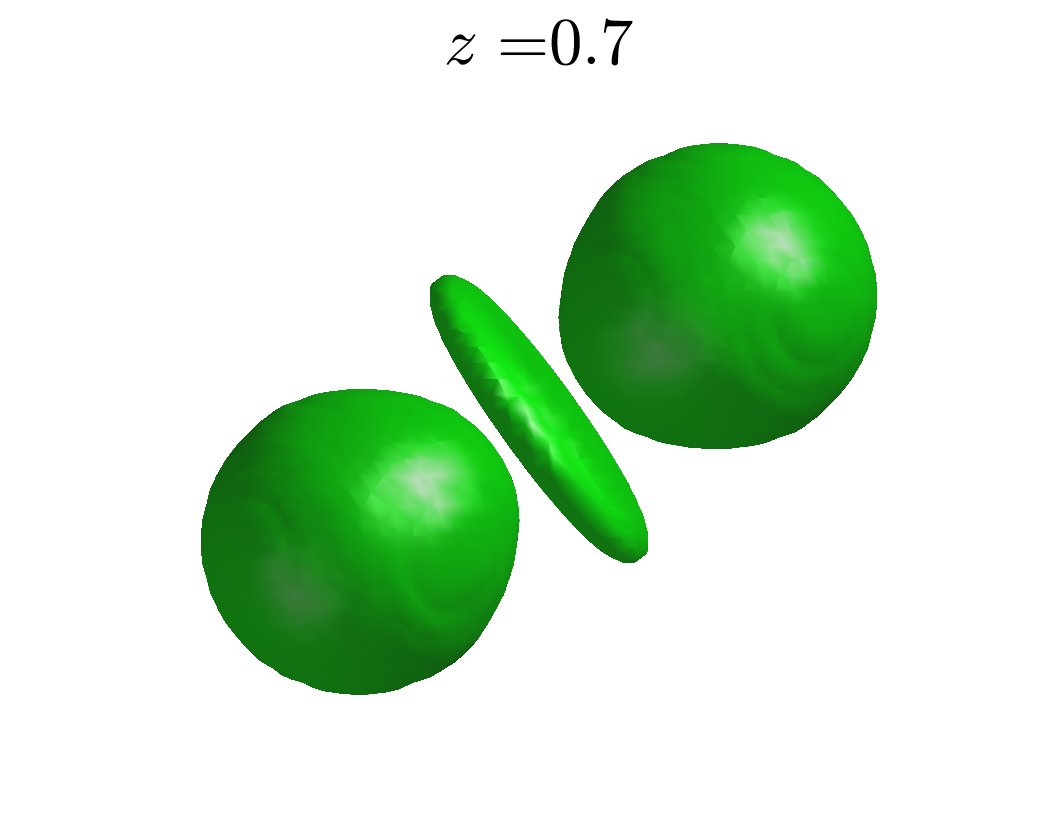}}
\subfigure[]{\label{co3}\includegraphics[height=4 cm, width=5 cm]{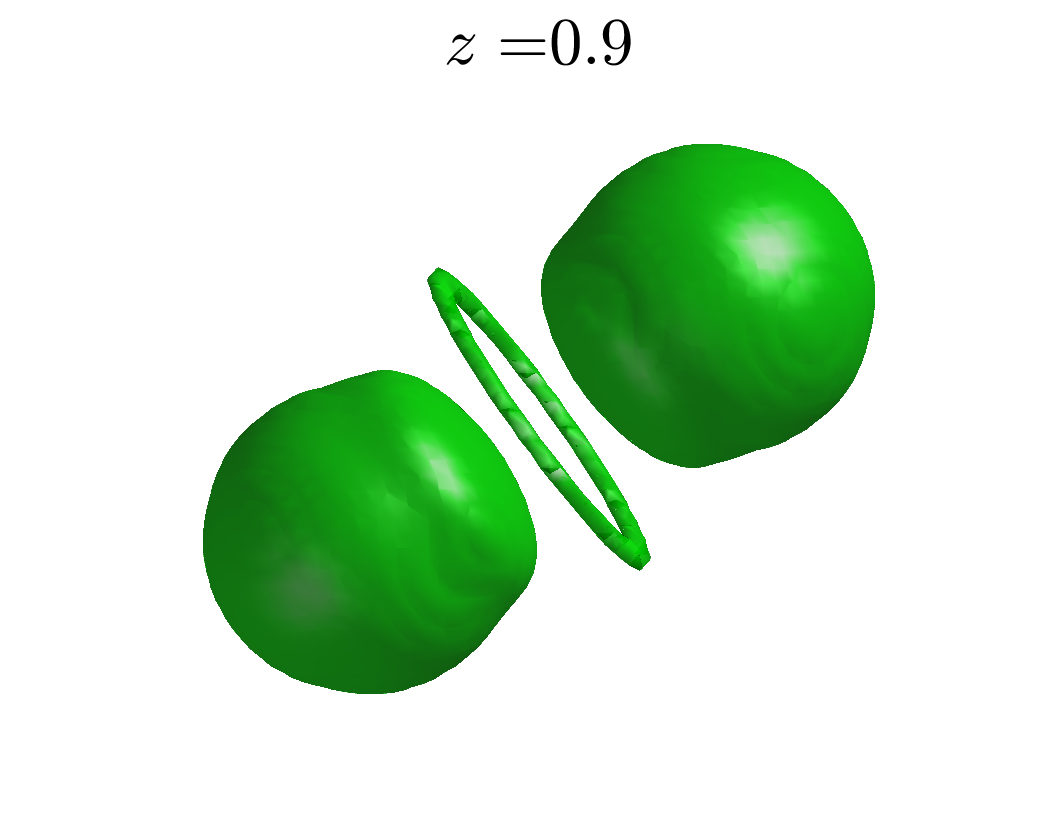}}
\subfigure[]{\label{co4}\includegraphics[height=4 cm, width=5 cm]{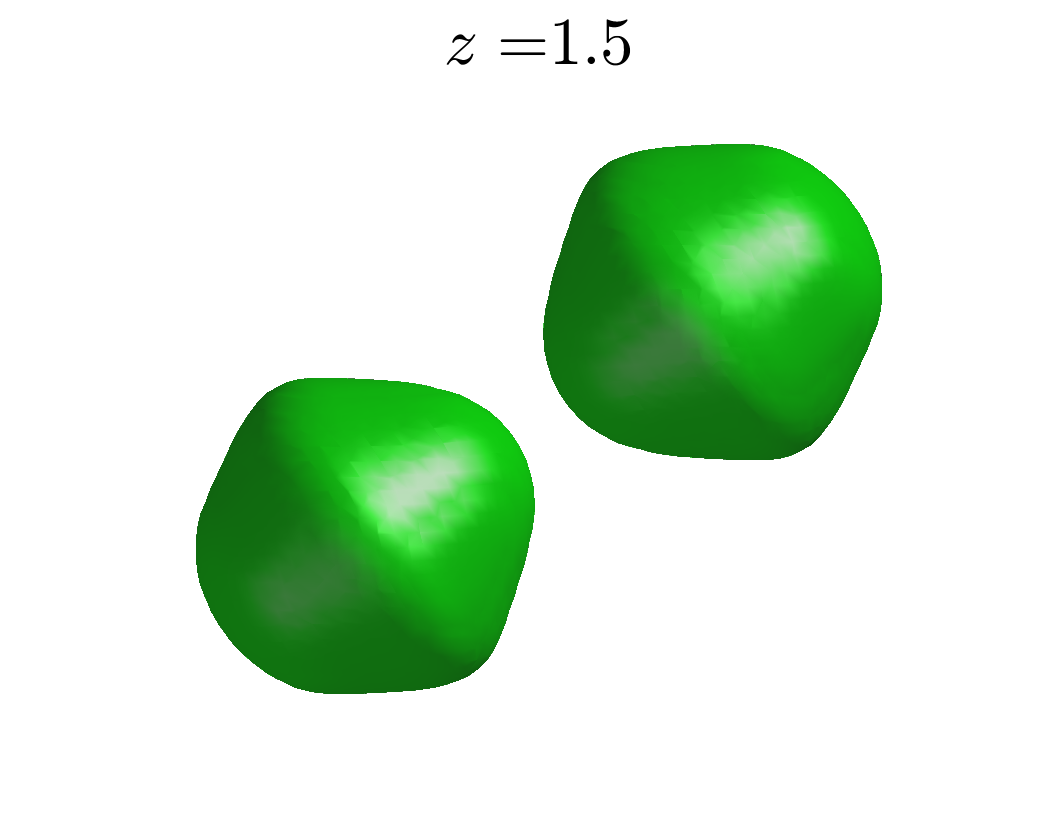}}
\subfigure[]{\label{co4}\includegraphics[height=4 cm, width=5 cm]{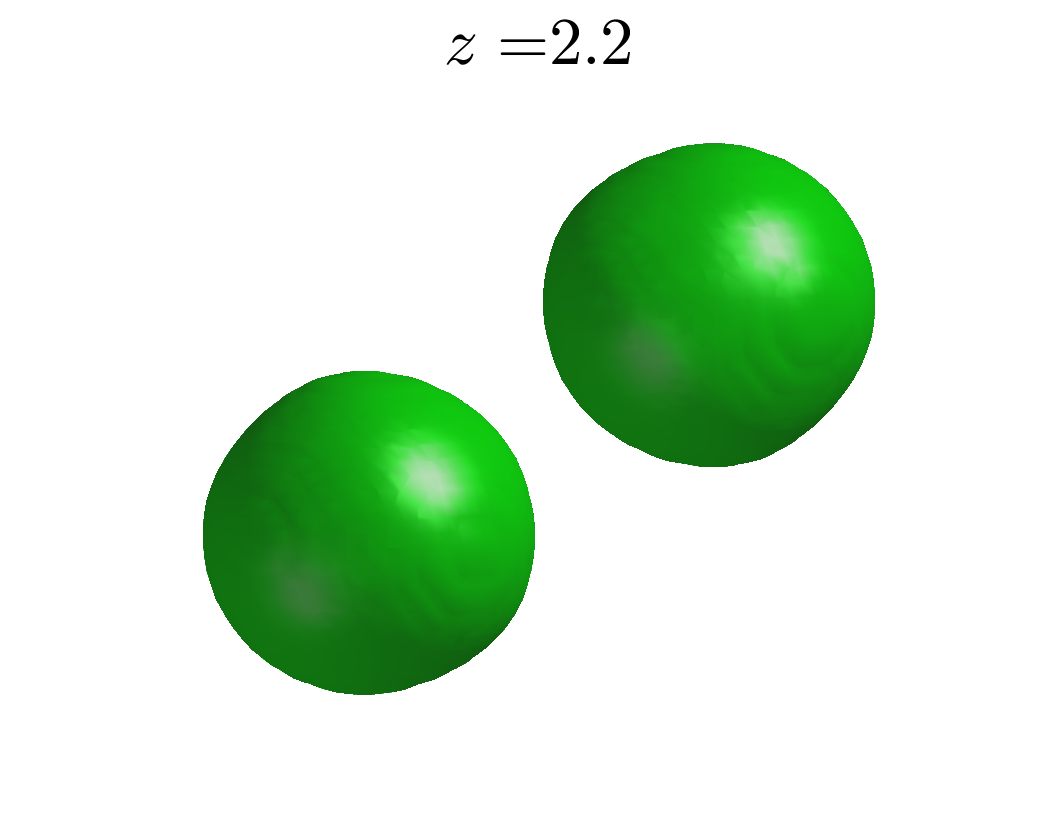}}
\caption{(Color online) Iso-surface plots showing the quasi-elastic collision dynamics of two light bullets in the negative index metamaterial with $P=1$, $\gamma$=-20 and $\vartheta$=15.}
\label{co}
\end{center}
\end{figure*}
In this section, we will uncover the interesting dynamics associated with light bullets in the negative index materials. It is well-known that in order to form the light bullets in optical media there should be a delicate balance among the nonlinearity, diffraction and dispersion. Hence the materials with appropriate values of system parameters such as dispersion, diffraction and nonlinearity are required to realize them experimentally. The parameters related to the conventional positive index materials are one of the great challenges to realize stable propagation of light bullets \cite{bu00, bu11}. As the metamaterial is engineered, it can get better off the limitations of the conventional materials. We study the parametric regions in which one can realize the stable propagation of the light bullets in the metamaterial waveguides.
 \par
We solve the governing equation (1) by the well-known split-step Fourier method implemented in the Scilab programming language.  The simulations have been performed on a 256x256x256 mesh grid with the initial condition (3) considered in the Lagrangian approach without the chirping effect. Also, other parameters are assigned as follows: $\Delta t=\Delta x=\Delta y=0.07$, $\Delta z=0.01$, and $\Phi=1$ throughout the study.
\par
As a result of optical beam collapse, the optical solitons in a medium with Kerr type optical nonlinearity are unstable in two and three dimensions. Figure \ref{a1} depicts iso-surface plots showing the propagation dynamics of a light bullet in the cubic nonlinear negative index metamaterial. The complete dynamics is given in supplemental material. Figure \ref{a1in} represents the input light bullet at $z$=0. As discussed in our analytical calculations, when the light bullet propagates in the cubic nonlinear metamaterial, it becomes unstable and cannot maintain its original shape. At $z$=0.15, the light bullet starts to exhibit breathing oscillations as shown in Fig. \ref{a12}. When it further propagates, the breathing oscillations grow and the bullet shape is distorted due to splitting (Fig. \ref{a14}) and ultimately it collapses into a chaotic state at a still larger propagation distance (Fig. \ref{a15}). This preponderance to collapse is due to the inability of Kerr nonlinearity to balance the inherent dispersive or diffractive effects present in the metamaterial waveguides. This ramification is as per the known Derrick's theorem, which states that the stationary localized solutions to certain nonlinear wave equations in higher dimensions are unstable \cite{der}.
\par
 \begin{figure*}
\begin{center}
\subfigure[]{\label{in11}\includegraphics[height=4 cm, width=5 cm]{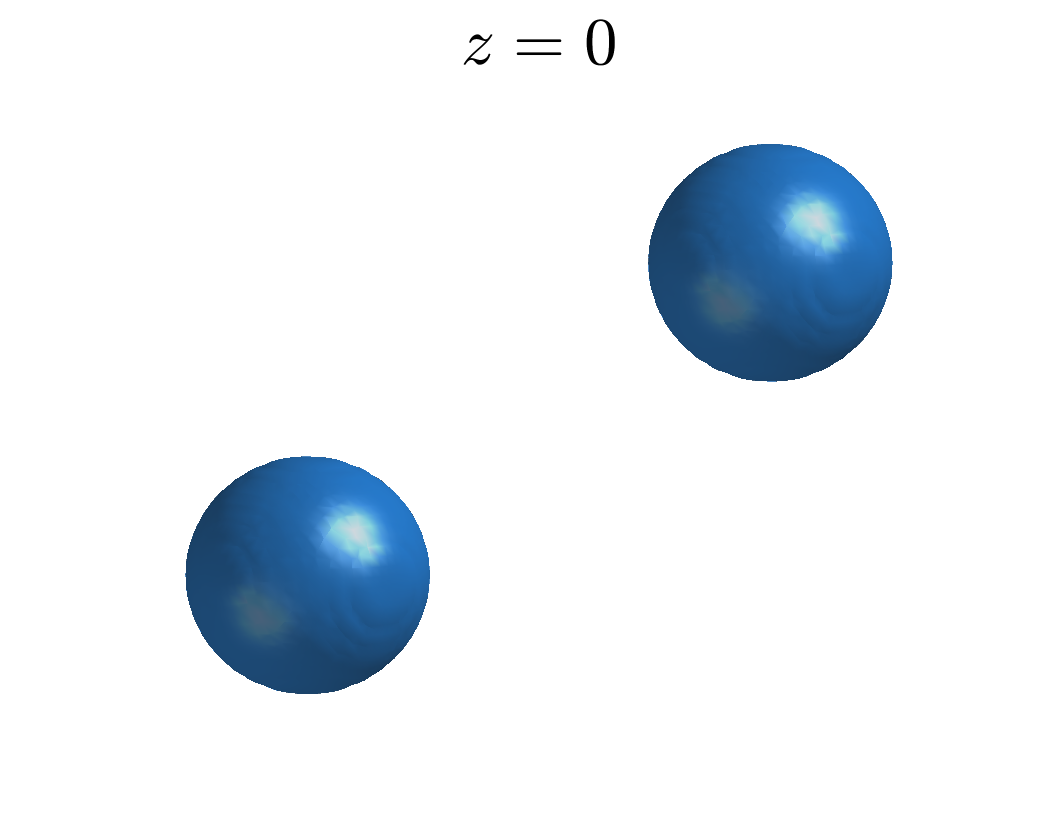}}
\subfigure[]{\label{in12}\includegraphics[height=4 cm, width=5 cm]{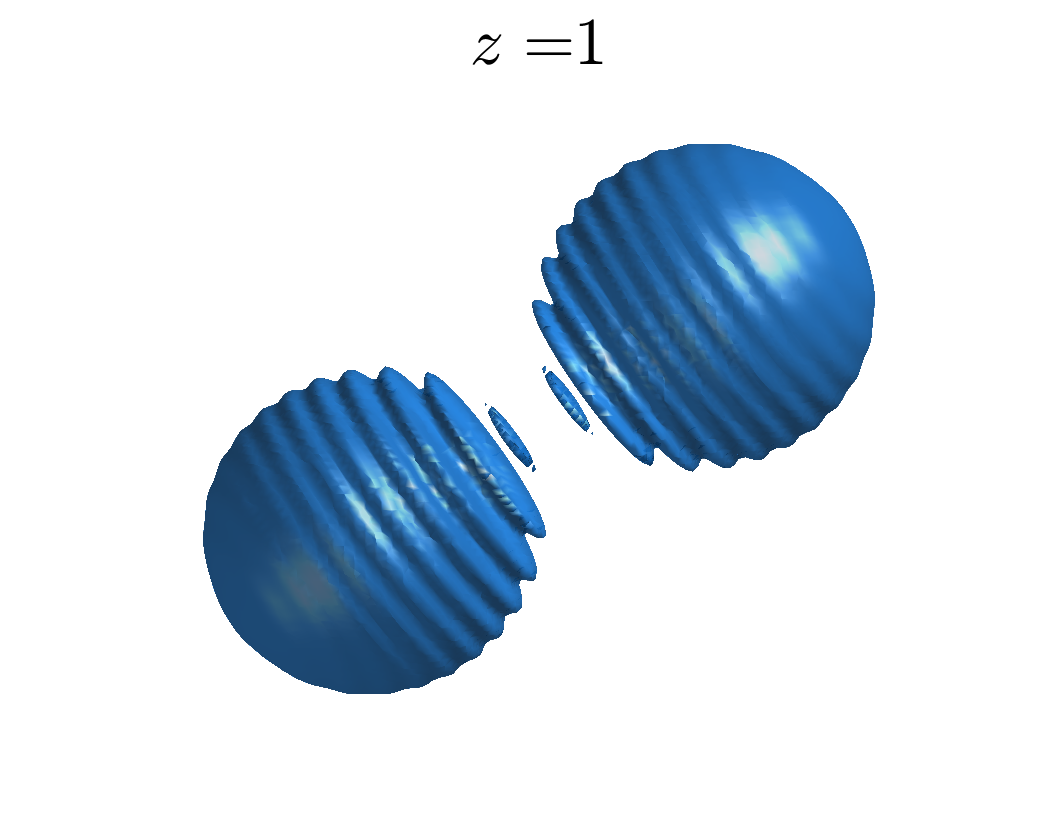}}
\subfigure[]{\label{in13}\includegraphics[height=4 cm, width=5 cm]{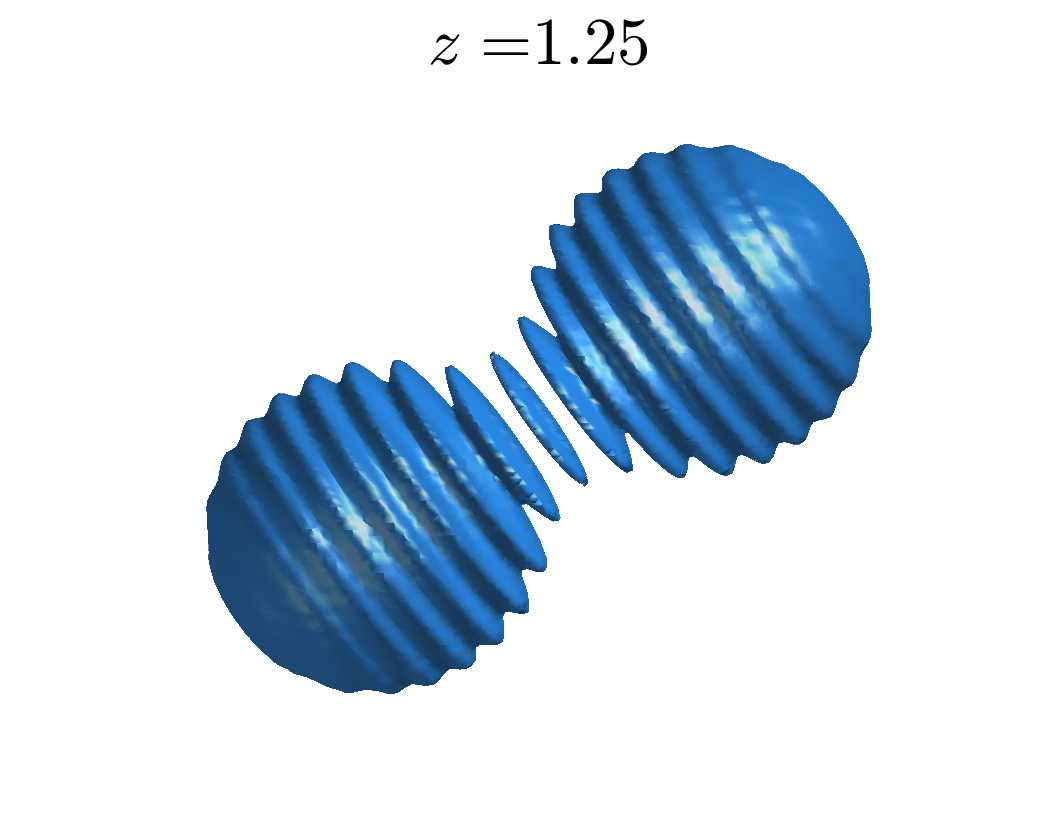}}
\subfigure[]{\label{in14}\includegraphics[height=4 cm, width=5 cm]{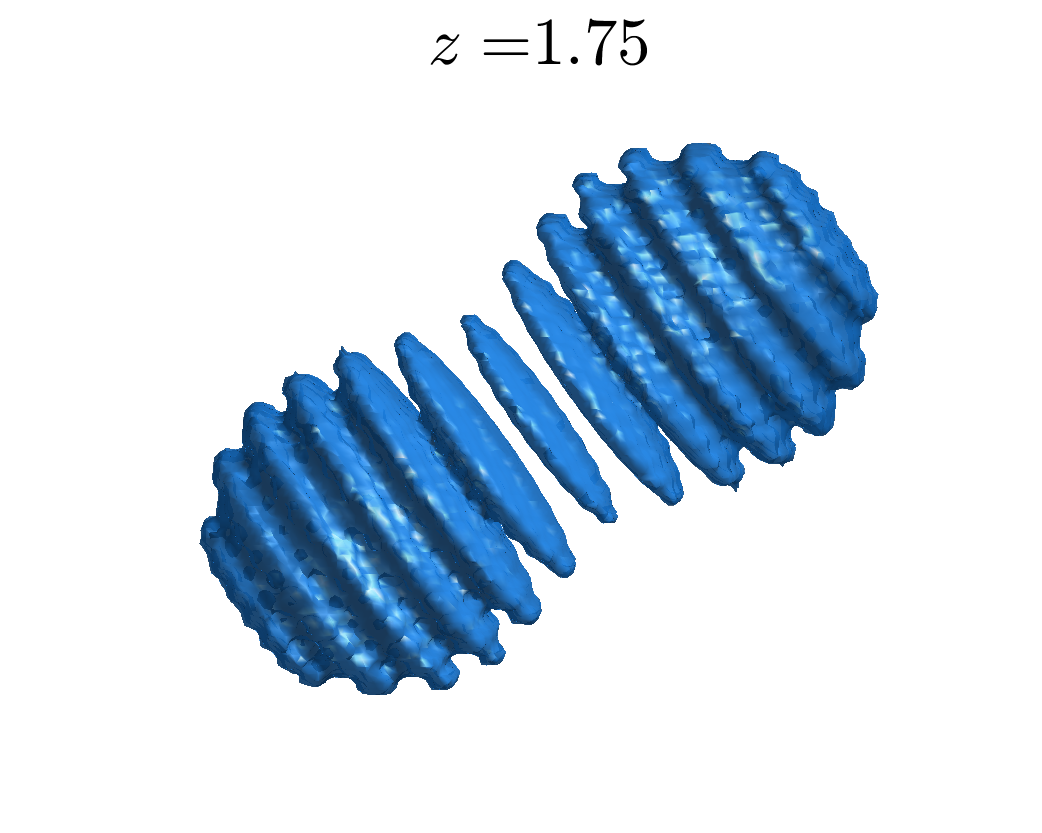}}
\subfigure[]{\label{in15}\includegraphics[height=4 cm, width=5 cm]{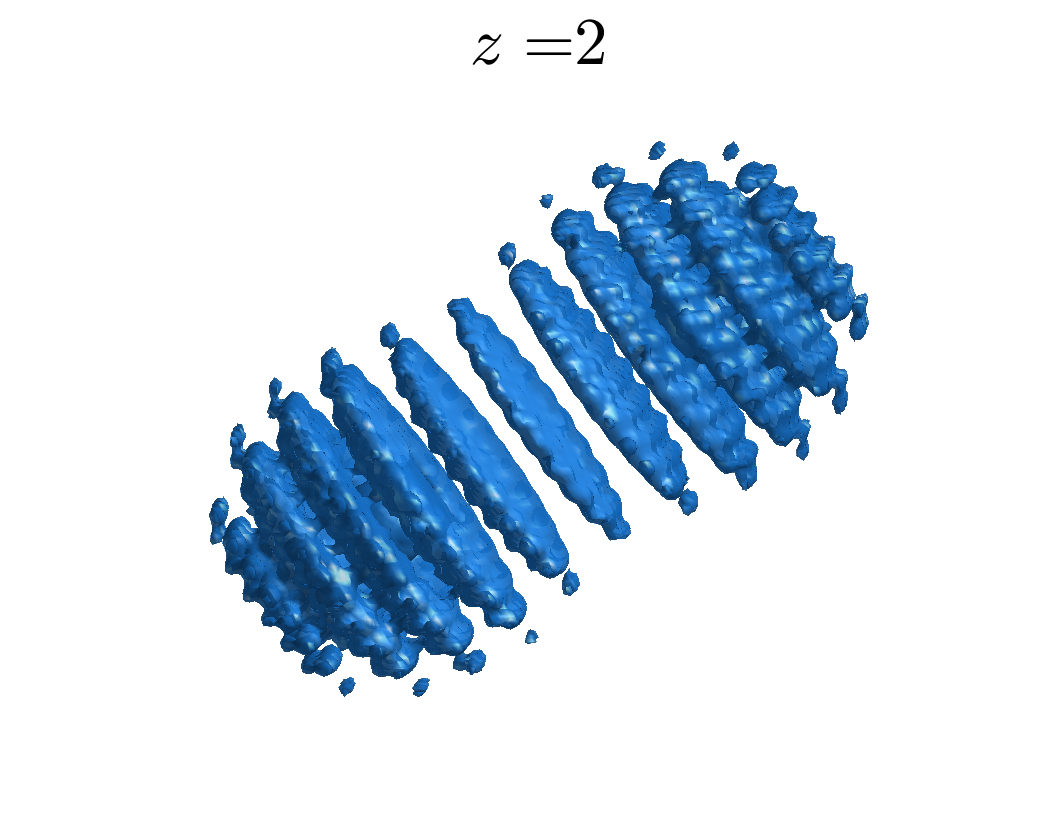}}
\subfigure[]{\label{in16}\includegraphics[height=4 cm, width=5 cm]{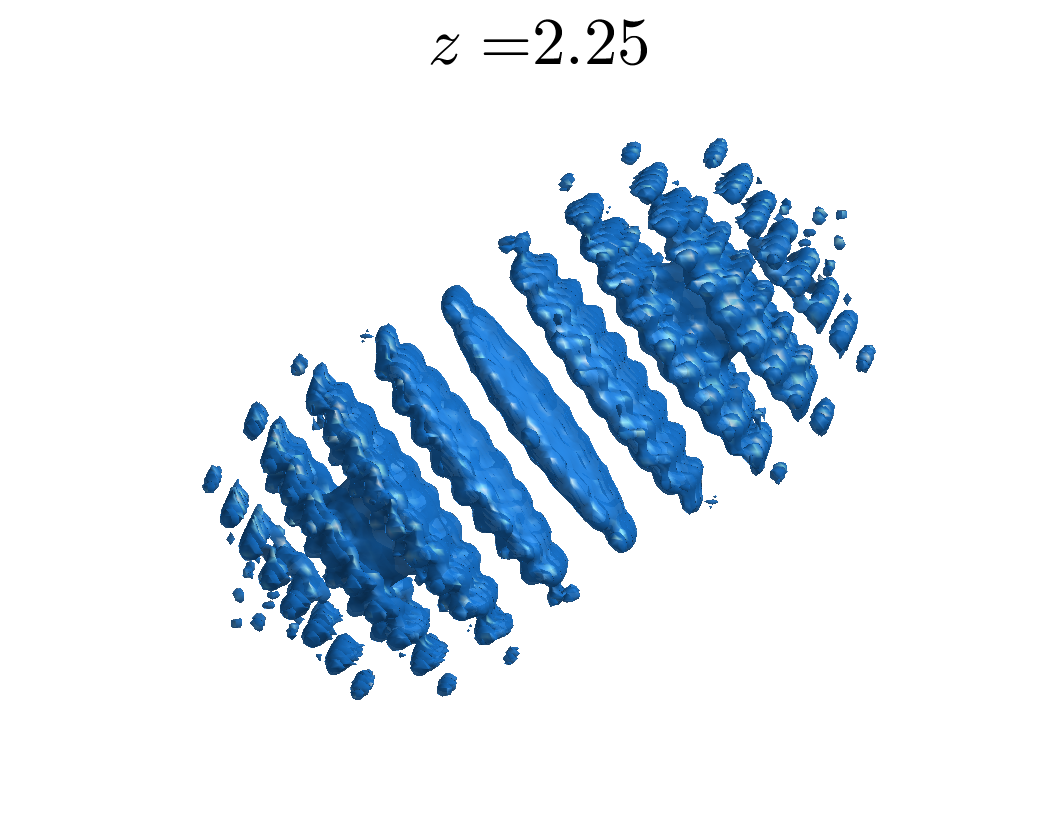}}
\caption{(Color online) Iso-surface plots portraying the inelastic collision dynamics of two light bullets in negative index metamaterial with $P=1$, $\gamma$=-30 and $\vartheta$=10.}
\label{coin}
\end{center}
\end{figure*}
The propagation of light bullets in the Kerr nonlinear metamaterial is unstable due to the insufficient balance between the nonlinearity and dispersive or diffractive effects. In order to achieve a proper balance and ensure the stable propagation of the light bullets, we consider the influence of quintic nonlinearity. Figure \ref{a2} depicts the evolution of light bullets in the negative index regime of the metamaterial with cubic and quintic nonlinearities. The complete collision dynamics is given in supplemental material. Here we consider the case of competing nonlinearities where the cubic nonlinearity is of the defocusing type and the quintic nonlinearity is of the focusing type. As compared to the cubic nonlinearity only case, the light bullet now shows more stability and propagates a larger distance without any splitting and distortion in the shape. This clearly indicates that the competing cubic and quintic nonlinearities present in the system provide stronger stability to the 3D light bullets. We note that we have already obtained the same result through the variational analysis. In the negative index regime of metamaterials, stable propagation of light bullets occurs in the normal dispersion regime balancing with the defocusing cubic nonlinearity and the focusing quintic nonlinearity whereas it is observed that in the anomalous dispersion regime with the focusing cubic nonlinearity and the defocusing quintic nonlinearity in the positive index regime of metamaterials.
\par
It is well known that the 1D solitons execute truly elastic collision and such solitons pass through each other without any deformation after the collision. Generally, the encounter between solitons is elastic, which indicates that the amplitude, shape and kinetic energy of both solitons are retained the same after interaction except for the phase shift \cite{co32}. In some special cases, such as a bright two-soliton solution of the integrable coupled nonlinear Schr\"{o}dinger equation \cite{ml01} and integrable N-coupled nonlinear Schr\"{o}dinger equations \cite{ml00} the shape-changing collision of solitons have also been reported. The shape-changing as a result of intensity redistribution among the interacting solitons has potential application to signal amplification and switching \cite{coml1}. Manakov equation can admit nondegenerate fundamental solitons which undergo collisions with and without energy redistribution \cite{sta1}.
 The collision between the two 3D vortex light bullets is quasi-elastic at large velocities \cite{co1, col0} and at medium velocities, the bullets may be destroyed
after the encounter. However, the collision between them is inelastic at small velocities and results in the formation of a breather after
encounter \cite{co1}. The collision between two light bullets with small velocities can lead to the formation of a bullet molecule \cite{co2}. Also, a study on the collision between two quantum balls reports that they can form a quantum-ball breather after collision \cite{co112}.

To test the solitonic dynamics of the 3D light bullets in the negative index metamaterials, we numerically investigate the collision between two light bullets. We consider the metamaterial waveguide, which exhibits normal dispersion and defocusing cubic and focusing quintic nonlinearities with $\gamma$=-20 and $\vartheta$=15. Figure \ref{co} depicts the collision dynamics of two 3D light bullets in the negative index metamaterial. The complete collision dynamics is given in supplemental material.  The bullets are initially separated by a distance of about 1.32 units as shown in Fig. \ref{coin} and they are set in motion. After the interaction, they form bullet molecule as depicted in Fig. \ref{co1}, then they partially oscillate due to splitting at certain propagation distances as depicted in Figs. (\ref{co2} and \ref{co3}). On further propagation, the light bullets regain their initial characteristics. From Fig. \ref{co}, it is easy to identify that the light bullet propagates without any change after a collision with another light bullet except perhaps for some phase shift. We believe that this quasi-elastic collision with no visible deformation of the 3D light bullets, in the defocusing cubic and focusing quintic nonlinear metamaterial waveguides is a remarkable result and which to the best of our knowledge has not been reported anywhere in the existing literature.
\par
We also report an inelastic collision between two light bullets in the nonlinear negative index material waveguides. Figure \ref{coin} depicts inelastic collision of one light bullet with another in the negative index metamaterial waveguide when $\gamma$=-30 and $\vartheta$=10. We have also provided the complete dynamics of inelastic collision in supplemental material. In this case, the bullets are initially separated by a distance of 2.5 units as shown in Fig. \ref{in11} and they are set in motion. During the propagation, they collide and form bullet molecules with severe oscillations as shown in Fig. \ref{in13}. As a result of the improper balancing between the linear and nonlinear parameters in the chosen range of parameters, the bullet molecules are in an excited state with a large amount of energy and, during further propagation, the oscillations of the light bullets grow and the bullets become filaments due to splitting into multiple pieces.
\section{Conclusion}
In conclusion, we have carried out a theoretical study on the dynamical behavior of the 3D light bullets in nonlinear metamaterial waveguides. Here we have brought out some unusual electromagnetic propagation properties that are not observed in the conventional media. It is found that in the negative index regime of the metamaterials stable propagation of light bullets can occur in the normal dispersion regime due to balancing between defocusing cubic nonlinearity and focusing quintic nonlinearity whereas it is observed in the case of anomalous dispersion with focusing cubic nonlinearity and defocusing quintic nonlinearity in the positive index regime. We also found that the light bullet propagates without any change except perhaps some phase shift after a collision with another light bullet in competing cubic and quintic nonlinear metamaterials. The improper balancing between the linear and nonlinear effects may lead to the formation of the bullet molecules in an excited state with a large amount of energy after the collision and during further propagation, the light bullets oscillations grow and the bullets become filaments. We would like to remark that invoking the role of Kerr nonlinearity has already been achieved in metamaterials way back in 2008 \cite{kivsharmeta} and so we hope that our theoretical study will aid in the experimental realization of stable dynamics of light bullets in metamaterials when the former would be embedded with semiconductor glasses like chalcogenide materials driven by the state-of the art technological advancements.
 \section{Acknowledgement}
 The work of A.K.S. is supported by the University Grants
Commission, Government of India, through a Dr. D. S.
Kothari Post Doctoral Fellowship in Sciences (Grant No.
F.4-2/2006 (BSR)/PH/18-19/0059).  A.G. acknowledges the support from University Grants Commission, Government of India, for providing a
Dr. D. S. Kothari Post Doctoral Fellowship in Sciences (Grant No. F.4-2/2006 (BSR)/PH/19-20/0025). M.L. is supported by a DST-SERB National Science Chair (Grant No.
NSC/2020/000029).
\section{Appendix}
In this Appendix we provide the general mathematical expressions of $P_{c1}$ (the critical power at which the equilibrium point at the potential energy curve exists in a cubic nonlinear medium) and $P_{c2}$ (the critical power required to support the formation of spatio-temporal soliton by a cubic and quintic nonlinear medium). If an  equilibrium point at the potential energy curve (Eq. \ref{Vqq}) exists, then $\frac{dV(a)}{da}=0$. In the case of cubic nonlinearity alone the above condition for the equilibrium points indicates
 \begin{eqnarray}
\frac{\gamma P}{3\sqrt{2}\pi^{\frac{3}{2}}a(z)^4}(sgn(n)-\frac{sgn(\beta_2)}{2})+\frac{4}{9a(z)^3} sgn(n) sgn(\beta_2)\nonumber \\
 -\frac{4sgn(n)^2}{9a(z)^3}-\frac{sgn(\beta_2)^2}{9a(z)^3}=0.
 \label{potc}
 \end{eqnarray}
 Consequently, the optical power to observe the equilibrium point in a cubic medium is given by
 \begin{eqnarray}
P_{c1}=\frac{2 \sqrt{2}\pi ^{3/2} a(z)}{3 \gamma} (2 sgn(n)-sgn(\beta_2)).
 \label{powe4}
 \end{eqnarray}
 Following the same procedure one can obtain the corresponding optical power in the cubic and quintic nonlinear medium, which is given by
 \begin{eqnarray}
 P_{c2}=-\frac{9 \sqrt{\frac{3}{2}}a(z)^3 \gamma \pi ^{3/2}}{16 \vartheta}\pm\nonumber \\ \frac{\sqrt{1458 \gamma^2 \pi ^3 a(z)^6-2304 \pi ^3 \sqrt{3} \vartheta a(z)^4( sgn(\beta_2) -2 sgn(n))}}{32 \sqrt{3} \vartheta}.
 \label{pow2}
 \end{eqnarray}
 For the stable equilibrium point $\frac{d^2V(a)}{da^2}>0$, for the unstable equilibrium point $\frac{d^2V(a)}{da^2}<0$. The stability of the equilibrium point depends on the nature of the medium and on the nonlinear parameters.

\end{document}